\documentclass[aps,twocolumn, english, showpacs, superscriptaddress, noeprint, prl, nofootinbib, nolinenumbers]{revtex4-2}
\usepackage{geometry}
\usepackage[T1]{fontenc}
\usepackage{xcolor}
\usepackage{babel}
\usepackage{units}
\usepackage{braket}
\usepackage{amsmath}
\usepackage{amssymb}
\usepackage{stmaryrd}
\usepackage{graphicx}
\usepackage{wasysym}
\usepackage{bbold}
\usepackage{bbm}
\usepackage{placeins}
\usepackage{soul}
\usepackage[colorlinks=true,citecolor=green,linkcolor =blue]{hyperref}
\usepackage[capitalise]{cleveref}
\usepackage[all]{hypcap} 
\usepackage{times}
\usepackage{txfonts}
\usepackage{orcidlink}

\makeatletter

\IfFileExists{lmodern.sty}{\usepackage{lmodern}}{}
\usepackage{hyperref}
\DeclareMathOperator{\Tr}{Tr}
\newcommand{\JQI}{Joint Quantum Institute, NIST and University of Maryland, College Park, Maryland 20742}
\newcommand{\QUICS}{Joint Center for Quantum Information and Computer Science, NIST and University of Maryland, College Park, Maryland 20742}
\newcommand{\UMDMCFP}{Department of Physics and Maryland Center for Fundamental Physics, University of Maryland, College Park, Maryland 20742}
\newcommand{\OXFORD}{The Rudolf Peierls Centre for Theoretical Physics, Oxford University, Oxford OX1 3NP, UK}
\newcommand{\LEUVEN}{Institute for Theoretical Physics, KU Leuven, Celestijnenlaan 200D, 3001 Leuven, Belgium}
\newcommand{\DUKE}{Duke Quantum Center, Departments of Physics and Electrical and Computer Engineering, Duke University, Durham, North Carolina 27708, USA}
\newcommand{\CALTECH}{Department of Physics and Institute for Quantum Information and Matter,
California Institute of Technology, Pasadena, California 91125, USA}
\newcommand{\CORNELL}{School of Applied and Engineering Physics, Cornell University, Ithaca, NY 14853.}
\newcommand{\QLAB}{National Quantum Laboratory (QLab),
University of Maryland, College Park, MD 20742 USA}

\makeatother

\begin{document}
\title{Observation of string-breaking dynamics in a quantum simulator}
\date{\today}
\author{Arinjoy De$^{*}~\orcidlink{0000-0001-9184-8434}$}
\affiliation{\JQI}
\affiliation{\DUKE}
\author{Alessio~Lerose$^{*}$
\orcidlink{0000-0003-1555-5327}}
\affiliation{\OXFORD}
\affiliation{\LEUVEN}
\author{De~Luo~\orcidlink{0000-0002-4842-0990}}
\affiliation{\DUKE}
\author{Federica~M.~Surace~\orcidlink{0000-0002-1545-5230}}
\affiliation{\CALTECH}
\author{Alexander~Schuckert~\orcidlink{0000-0002-9969-7391}}
\affiliation{\JQI}
\affiliation{\QUICS}
\author{Elizabeth~R.~Bennewitz}
\affiliation{\JQI}
\affiliation{\QUICS}
\author{Brayden~Ware}
\affiliation{\JQI}
\affiliation{\QUICS}
\author{William~Morong$^{\dagger}$}
\affiliation{\JQI}
\affiliation{\QUICS}
\author{Kate~S.~Collins}
\affiliation{\JQI}
\affiliation{\QUICS}
\author{Zohreh~Davoudi~\orcidlink{0000-0002-7288-2810}}
\affiliation{\UMDMCFP}
\affiliation{\QUICS}
\affiliation{\QLAB}
\author{Alexey~V.~Gorshkov}
\affiliation{\JQI}
\affiliation{\QUICS}
\author{Or~Katz~\orcidlink{0000-0001-7634-1993}}
\affiliation{\DUKE}
\affiliation{\CORNELL}
\author{Christopher~Monroe}
\affiliation{\DUKE}
\def\thefootnote{*}\footnotetext{These authors contributed equally to this work. \\Email: alessio.lerose@kuleuven.be, arinjoy.de@duke.edu}
\def\thefootnote{$\dagger$}\footnotetext{Current address: AWS Center for Quantum Computing, Pasadena, California 91125, USA. Work done prior to joining AWS.}

\maketitle

\textbf{The spontaneous formation of particle pairs is one of the most intriguing phenomena in nature. A canonical example is when the potential energy between two elementary particles increases with their separation, as if the particles were confined to one another by a tense string. When the separation exceeds a critical value, new particle pairs can form, which causes the string to break. String-breaking dynamics in quantum chromodynamics, the theory of the strong force, play a fundamental role in high-energy particle collisions and the evolution of matter in the early universe. 
Simulating the evolution of strings and the formation of composite particles, or hadrons, is a grand challenge of modern physics. Quantum simulators are well suited to study dynamical processes. They are, therefore, expected to provide an edge over conventional methods based on classical computing.
However, the  experimental capabilities required to simulate the string-breaking phenomenon have not yet been demonstrated, even for simpler prototypical models of the strong force. 
In this work, we probe, for the first time on a quantum simulator, the spatiotemporally-resolved dynamics of string breaking in a $(1+1)$-dimensional $\mathbb{Z}_2$ lattice gauge theory, a theory that also exhibits confinement of charges.
We employ a fully programmable trapped-ion quantum simulator, and emulate the effects of external static charges and strings via site-dependent control of magnetic fields, using a dual array of tightly focused laser beams targeting individual ions. 
First, we study the effect of confinement on the evolution of isolated charges. We find that these charges freely spread in the absence of string tension, but exhibit localized coherent oscillations as the string tension is increased.
Then, we observe and characterize the breaking dynamics of a string initially stretched between two static charges, following an abrupt increase of the string tension. We find that charge pairs appear near the string edges and then spread out into the bulk, thereby identifying a route to dynamical string breaking that is distinct from the conventional Schwinger mechanism.
This work, therefore, demonstrates that analog quantum simulators have achieved the control needed to uncover features of string-breaking dynamics, which may ultimately be relevant to nuclear and high-energy physics.}

Atomic nuclei are described at a fundamental level by quantum chromodynamics of quarks and gluons. 
Quarks and gluons, which carry a net \emph{color} charge, are not found in isolation, but only in colorless agglomerates known as hadrons (e.g., protons and neutrons), tied together in
an effect referred to as color-charge \emph{confinement}~\cite{wilson1974confinement,greensite2011introduction}.
This phenomenon is commonly conceptualized via a paradigmatic thought experiment, whereby a pair of probe charges are pulled apart. 
In nature, pulling apart an electron and a positron to an infinite distance requires a finite electrostatic energy. 
By contrast, in quantum chromodynamics, the potential energy between a quark and an antiquark increases indefinitely with distance, stretching a gluon string that binds them together. As the separation increases, the energy eventually becomes larger than twice the quark mass, driving the creation of new quark-antiquark pairs, thus breaking the string. String-fragmentation dynamics are conjectured to govern the hadronization processes in relativistic heavy-ion collisions~\cite{ANDERSSON198331,Lund} and in the cooling of the expanding universe after the Big Bang~\cite{peebles1993principles,kolb2018early}. However, such phenomena remain poorly understood at a fundamental level. While classical simulations based on sampling methods have  successfully computed the static string breaking in quantum chromodynamics~\cite{bali2005observation}, they are severely inefficient in simulating string dynamics~\cite{gattringer2016approaches}. 
Other classical methods, such as tensor networks, may be more suited to study this problem~\cite{hebenstreit2013real,hebenstreit2013simulating,kuhn2015non,pichler2016real,buyens2017real,Verdel19_ResonantSB,Magnifico2020realtimedynamics,verdel2023dynamical}. However, such methods have yet to be developed for, and applied to, the theory of the strong force~\cite{banuls2020simulating,meurice2022tensor,magnifico2024tensor}, and they will likely be inefficient for high-energy processes with abundant entanglement generation.  

\begin{figure*}[ht]
    \centering
    \includegraphics{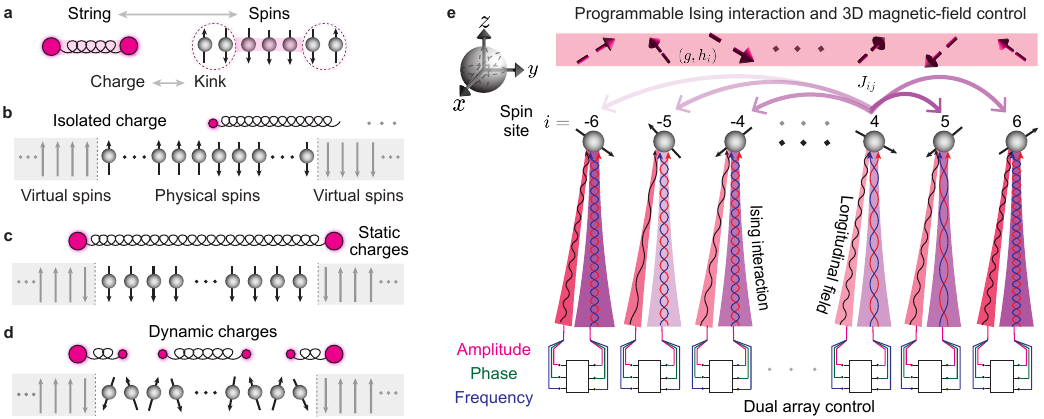}
    \caption{\textbf{Charge and string dynamics with a trapped-ion quantum simulator.} 
    \textbf{a.} Mapping between the degrees of freedom of a $(1+1)$-dimensional $\mathbb{Z}_2$ lattice gauge theory and an Ising spin model implemented by our simulator. Charges in the gauge theory are represented by kinks in the spin model, and strings are represented by down-spin domains.
    \textbf{b.} The initial state prepared for the simulation protocol of \cref{fig4:localized_charge}, which implements the dynamics of an isolated charge under Hamiltonian evolution. 
    The virtual chains (gray-shaded regions) realize a semi-infinite static string stretching to the right and a semi-infinite static vacuum on the left, enforcing a net dynamical charge in the physical region. \textbf{c.} The initial state prepared for the simulation protocol of \cref{fig3: static_charge}. The static-spin configurations at the edges of the two semi-infinite virtual chains realize a pair of static charges. 
    \textbf{d.} Sketch of a spin configuration representing a state with dynamically generated charge pairs. Such states can form during the simulation protocol of
    \cref{fig3: static_charge}, which implements far-from-equilibrium string-breaking dynamics.
    \textbf{e.} Schematic of the trapped-ion apparatus, showcasing two arrays of tightly focused laser beams that can be independently controlled. One of the arrays (purple) generates long-range spin-spin Ising interactions. The second array of beams (red) implements a site-dependent magnetic field. Each of the beam amplitudes (different shades) is controlled to generate a uniform interaction profile across the ion chain as well as a site-dependent local magnetic field (see Extended Data Figs.~\ref{fig:Jij_calibration} and \ref{fig:Bx_calibration}). Frequency and phase control of each beam (wiggly lines inside the beam) enables full tunability of the magnetic-field direction along any axis of the Bloch sphere $\boldsymbol{x, y, z}$ (dashed arrows on top). 
    }
    \label{fig_cartoon}
\end{figure*}

By employing quantum simulators---in analog, digital, or hybrid modes---the same dynamical processes can be simulated efficiently, i.e., their simulation requires computational resources that scale only polynomially with system size and evolution time. As a result, quantum technologies may present a promising path toward simulating quantum chromodynamics~\cite{byrnes2006simulating,wiese2013ultracold,ZoharReview,dalmonte2016lattice,banuls2020simulating,klco2022standard,aidelsburger2022cold,bauer2023quantumsimulation,bauer2023quantum,PRXQuantum.5.037001}.
While a quantum-chromodynamics simulator remains beyond current capabilities, existing platforms have started to investigate charge confinement in simplified lower-dimensional models.
These experiments have been realized using trapped ions~\cite{martinez2016real,nguyen2022digital,tan2021domain,farrell2023preparations,davoudi2024scattering,meth2023simulating}, neutral-atom tweezer arrays~\cite{bernien2017probing,SuraceRydberg}, optical lattices~\cite{schweizer2019floquet,yang2020observation,zhou2022thermalization}, and superconducting-qubit processors~\cite{klco2018quantum,mildenberger2022probing,charles2024simulating,farrell2024quantum}.
However, simulating phenomena such as hadron collisions~\cite{Surace_2021,karpov2022spatiotemporal,bennewitz2024simulating,su2024cold} or string-fragmentation dynamics demands a higher level of control precision and programmability than that achieved to date.

In this work, we develop, and experimentally implement, an efficient strategy to study the dynamics of confined charges and strings using an analog trapped-ion quantum simulator. Precise control over individual ions allows us to emulate the effects of virtual static environments, that we use to induce a net charge or a string.
A rapid change of  Hamiltonian parameters then drives non-equilibrium charge or string dynamics, which we observe with full spatiotemporal resolution. Our results push the boundaries of quantum simulation, providing new insights into quantum many-body dynamics, with potential relevance to nuclear and high-energy physics.

\medskip

Confinement, a hallmark feature of quantum chromodynamics in $(3+1)$ dimensions, can also be studied using $(1+1)$-dimensional gauge theories~\cite{SchwingerModel,tHooftModel,McCoyWuConfinement}. In this work, we study dynamics of a $\mathbb{Z}_2$ lattice gauge theory consisting of fermionic matter residing on the sites of a one-dimensional lattice, coupled to $\mathbb{Z}_2$  gauge fields residing on the bonds (links) between lattice sites. The fermionic charge at each site is constrained by the electric fields at adjacent lattice bonds via Gauss's law. 
Alternatively, Gauss's law can be used to eliminate the fermions from the description of dynamics~\cite{ZoharRemovingFermions}. The resulting fully bosonic dual theory, as shown in Methods and illustrated in Extended Data~\cref{fig_mapping}, can be mapped to a one-dimensional quantum Ising chain with the Hamiltonian~\cite{balian1975gauge,zhang2018quantum,lerose2020quasilocalized,borla2020gauging}
\begin{equation}
\label{eq_H}
    H = - \sum_{i<j}J_{i,j} {\sigma}_i^{z} {\sigma}_j^{z} - \sum_{i}h{\sigma}_i^{z} - \sum_{i}g {\sigma}_i^{x}.
\end{equation}
Here, $\sigma^\alpha_i$ with $\alpha=x,z$ are the Pauli matrices at spin site $i$, and $h\ge 0$ and $g \ge 0$ represent the strength of the longitudinal and transverse components of the magnetic field, respectively. Additionally, $J_{i,j}>0$ is the Ising-interaction couplings between spins $i$ and $j$.

A charge in the original formulation is encoded as a kink in the dual formulation, where a kink is a pair of anti-aligned spins $\uparrow\downarrow$ or $\downarrow\uparrow$. 
The charge density is quantified by $q_i=\langle1-\sigma^z_{i-1}\sigma^z_{i}\rangle/2$.
On the other hand, the local longitudinal spin polarization encodes the electric field, denoted $\epsilon_i=\langle\sigma^z_i\rangle$. 
An electric-field string is associated with a down-spin domain $ \downarrow\cdots\downarrow$ . Figure~\ref{fig_cartoon}a depicts degrees of freedom in the original lattice gauge theory and the dual spin theory. The couplings $J_{i,j}$ control the particle mass and short-range interactions. 
The longitudinal field $h$ generates a \emph{string tension}, i.e., the energy stored in the string per unit length.  
The transverse field introduces quantum fluctuations that couple charge and string dynamics, with coupling strength $g$. The tendency of strings to break dynamically and form new charge pairs is enhanced through both an increased coupling strength and an increased string tension.

Using a quantum simulator to study the charge and string dynamics induced by the Hamiltonian in Eq.~\eqref{eq_H} poses two main challenges. First, experiments can only simulate finite systems. To make the best use of quantum resources, we analyze a simpler system comprised of infinitely long arrays of static spins to the left and right of a finite domain of dynamical spins. While only the dynamical spins are encoded in the quantum simulator, an additional site-dependent longitudinal field, $\Delta h_i$, can be applied to the dynamical spins to emulate interactions with the static spins (see  Methods). This scheme enables the realization of external static strings or charges, as depicted in Fig.~\ref{fig_cartoon}b,c, and alleviates the undesired effects of hard boundaries. It, nonetheless, requires site-dependent control of $h$ in the experiment. This requirement is related to the second experimental challenge, which is the need for simultaneous implementation of all the Hamiltonian terms with programmable strengths.

We address these challenges by using a trapped-ion quantum simulator with enhanced control. We encode the spins into two internal levels of $^{171}$Yb$^+$ ions and implement the transverse-field Ising model [the first and third terms in Eq.~\eqref{eq_H}] with a linear array of $L=13$ spins addressed via bichromatic laser beams. 
These beams off-resonantly couple to the collective normal modes of ion motion~\cite{Feng2023,katz2024observing,schuckert2023}. Simultaneous control of the amplitude, phase, and frequency of each beam in the array allows us to realize the uniform field $g$, as well as to program the interaction profile $J_{i,j}$ to be translationally invariant and exponentially decaying with distance, $J_{i,j}= J e^{-\beta (|i-j|-1)}$ with $\beta \simeq 0.78$ (see Methods). 
To realize the site-dependent longitudinal field [the second term in Eq.~\eqref{eq_H} with $h \to h_i = h+\Delta h_i$], we apply a second array of tightly focused beams that drive carrier Raman transitions, with independent control over the amplitude, phase, and frequency of each beam, as shown in Fig.~\ref{fig_cartoon}e. This setup enables, for the first time in trapped-ion experiments, simultaneous and independent control of the local magnetic-field vector at each spin site, in addition to programmable long-range spin-spin interactions. 

\medskip

\begin{figure*}[ht]
\includegraphics{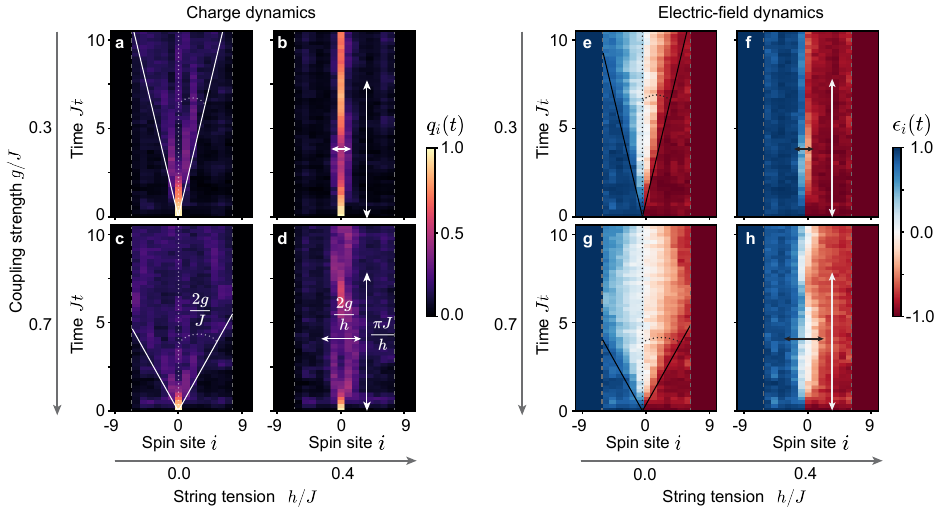}
\caption{\textbf{Non-equilibrium charge dynamics.} 
Time evolution of a perfectly localized charge at the center of the physical-spin chain (Fig.~\ref{fig_cartoon}b) in the presence of a range of coupling-strength and string-tension values.  
Panels \textbf{a-d} depict evolution of the charge density $q_i = \langle1-\sigma^z_{i-1}\sigma^z_{i}\rangle/2$, while panels \textbf{e-h} depict the evolution of the  electric field $\epsilon_i=\langle\sigma^z_i\rangle$. Note that initially, the charge distribution is $q_i=\delta_{i,0}$, the region to the left of the charge is the classical vacuum state with the electric field $\epsilon_i=\langle\sigma^z_i\rangle=1$, and the region to the right of the charge is a (semi-infinite) classical string state with $\langle\sigma^z_i\rangle=-1$.   
The superimposed lines and arrows correspond to a single-charge approximation of dynamics: lines $i=\pm v_{\rm{max}} t$ in panels \textbf{a,c,e,g}, with $v_{\rm{max}}=2g$, highlight the maximum speed of propagation of the charge, and arrows in panels \textbf{b,d,f,h} highlight amplitude ($2g/h$) and period ($\pi/h$) of the coherent oscillations.
Dashed vertical lines mark the boundaries between the simulated physical region and the virtual static environments. Further exploration of the parameter space for both charge and electric-field dynamics, along with their comparisons to numerical simulations, is presented in Extended Data~\cref{fig:domain_wall_sim_charge,fig:domain_wall_sim_field}, respectively. 
}\label{fig4:localized_charge}
\end{figure*}

\begin{figure*}
\includegraphics[width=\textwidth]{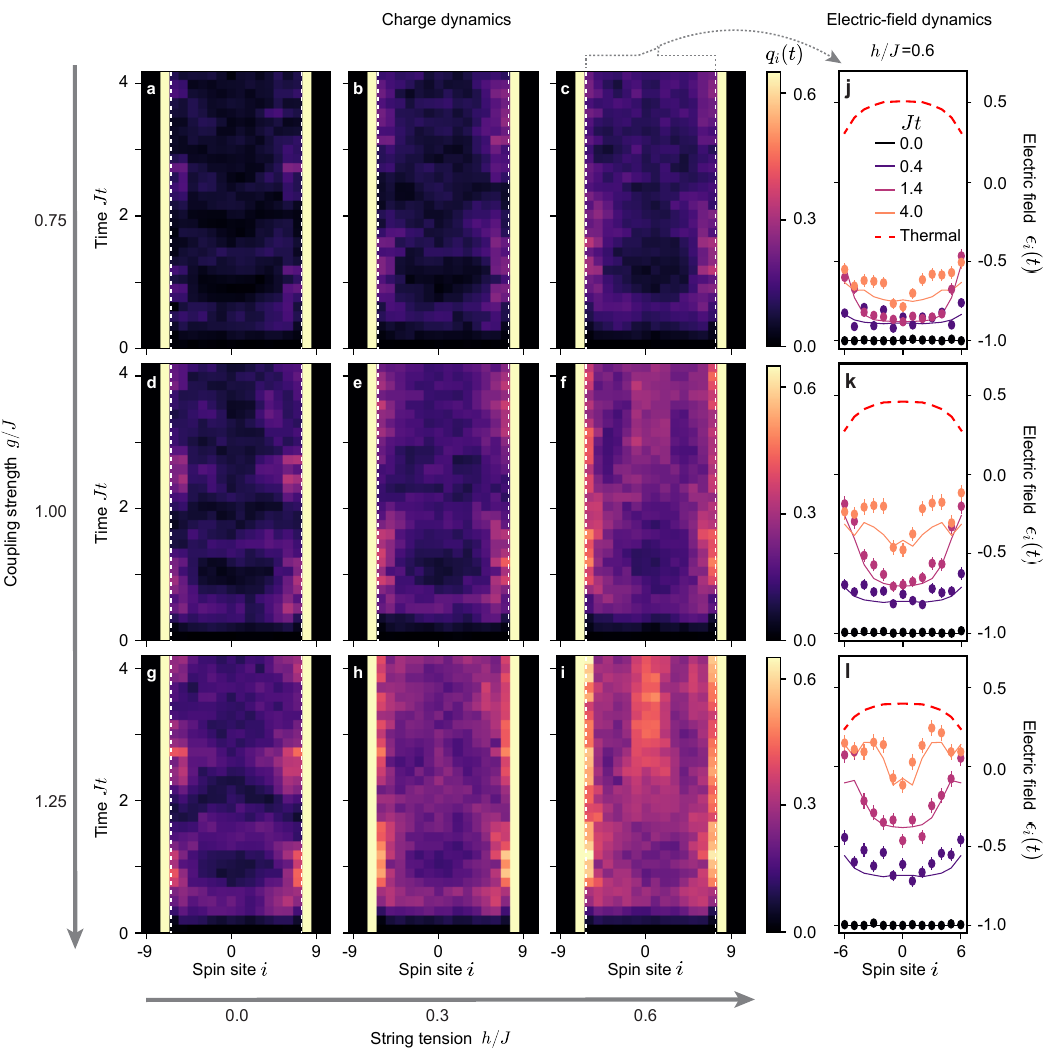}
\caption{\textbf{
Non-equilibrium string dynamics.} 
Time evolution of a classical string state (\cref{fig_cartoon}c) with a pair of external static charges.
Panels \textbf{a-i} depict evolution of the charge density $q_i=\langle (1-\sigma^z_{i-1}\sigma^z_{i})/2 \rangle$, with coupling strength ($g/J$) and string tension ($h/J$) varying across the grid. The initial charge distribution is $q_i=\delta_{i,-7}+\delta_{i,8}$, with the yellow bars denoting the position of static charges at the boundary.
The corresponding numerical results are shown in Extended Data Fig.~\ref{fig:static_charge_sim}. 
Panels \textbf{j-l} depict the electric field $\epsilon_i=\langle \sigma^z_i \rangle$, measured at different times for $h=0.6J$ (dots with error bars), together with the corresponding numerically simulated profiles (solid lines), and contrasted to the expected profiles in thermal equilibrium (red dashed lines). The electric-field dynamics for all the parameters and the corresponding numerical simulations are reported in the Extended Data~\cref{fig:static_charge_efield,fig:static_charge_efield_sim}, respectively.
All experimental measurements were averaged over 300 repetitions, and error bars represent statistical fluctuations around the mean value. 
} \label{fig3: static_charge}
\end{figure*}

In preparation for studying string dynamics, we first focus on the evolution of an isolated charge.
While an isolated quark cannot be observed in particle colliders, in our quantum simulator an isolated charge is realized by emulating the effect of a semi-infinite static string extending to the right and a semi-infinite vacuum extending to the left of the simulated physical region, as illustrated in \cref{fig_cartoon}b.  The expression of the corresponding site-dependent field, $\Delta h_i^{\text{(charge)}}$, is provided in Methods. 
We study the spreading dynamics of the string endpoint, which may precede and affect bulk string breaking. The interplay between these two dynamical processes has been first theoretically addressed in Refs.~\cite{Verdel19_ResonantSB,verdel2023dynamical} for the specific case of resonant string breaking.  

In Fig.~\ref{fig4:localized_charge}a-d, we present the evolution of the charge distribution governed by several Hamiltonians with non-vanishing interactions, starting from the classical (i.e.\ fluctuation-free)  initial state shown in Fig.~\ref{fig_cartoon}b, with charge localized at the center of the simulated region. The corresponding electric-field spatiotemporal dynamics are shown in Fig.~\ref{fig4:localized_charge}e-h. In the absence of string tension ($h=0$),  charge spreads ballistically across the physical lattice until it hits the static boundaries, as shown in Fig.~\ref{fig4:localized_charge}a,c,e,g. The measured velocity matches well with the value $v_{\rm{max}}=2g$, calculated by retaining charge-hopping processes and neglecting charge-pair creation processes, which is justified for small $g/J$. As the string tension is increased, we observe a Wannier-Stark localization phenomenon~\cite{WannierStarkLocalization}, shown in Fig.~\ref{fig4:localized_charge}b,d,f,h, which can be understood within the same single-charge approximation~\cite{lerose2019quasilocalized}. 
The string tension imparts a constant acceleration to the charge. 
In continuum space, a string can pull a charge indefinitely far, as its momentum can grow unbounded. However, on a lattice, the momentum is limited to a finite range. A constant acceleration, therefore, generates a periodic variation of momentum in time. This gives rise to coherent oscillations of the charge around its initial position, akin to Bloch oscillations, originally predicted for an electron moving in a defect-free crystal lattice under a constant electric field~\cite{bloch1929quantenmechanik}. This single-charge description of dynamics  becomes quantitatively accurate for small $g/J$ and $h/J$, where the spatial amplitude and temporal period of the oscillations (in units of inverse $J$) are predicted to be $2g/h$ and $\pi J/h$, respectively~\cite{MazzaTransport,lerose2019quasilocalized}. We find good agreement between these theoretical predictions and the data, even for non-perturbatively large $g/J$ and $h/J$, and between the experiment and numerical simulations, as shown in the Extended Data Figs.~\ref{fig:domain_wall_sim_charge} and~\ref{fig:domain_wall_sim_field}. 
Overall, \cref{fig4:localized_charge} demonstrates that the main qualitative effect of charge confinement is to halt charge spreading dynamics, thereby localizing string endpoints~\cite{lerose2020quasilocalized,MazzaTransport,Verdel19_ResonantSB}. 

 \medskip

In order to study the non-equilibrium evolution of strings, it is convenient to disentangle the spatial dynamics of string endpoints, observed in Fig.~\ref{fig4:localized_charge}, from the genuine charge-pair-creation processes arising from string breaking. To do so, we will consider a string delimited by static charges and surrounded by a static vacuum, as illustrated in \cref{fig_cartoon}c. To accurately realize this configuration experimentally, we include a set of site-dependent longitudinal-field values $\Delta h_i^{\text{(string)}}$ in the simulated physical region. This field emulates the coupling with the two semi-infinite virtual chains shown in \cref{fig_cartoon}c.  
Similar to our simulation protocol in \cref{fig4:localized_charge}, we prepare the physical region at time $t=0$ in the classical (i.e.\ fluctuation-free) string state---which is the ground state with vanishing string tension and coupling strength ($g=h=0$)---and abruptly increase both $g$ and $h$. The observed out-of-equilibrium string dynamics are shown in \cref{fig3: static_charge}. 
We have explored a wide range of parameter values encompassing stronger abrupt changes in the coupling strength and the string tension compared to our isolated-charge experiment in \cref{fig4:localized_charge}. 
These stronger parameter values correspond to an enhanced probability of charge-pair creation in the ensuing time evolution.  

The abrupt changes in the Hamiltonian parameters inject a finite energy density into the system. The strong many-body interactions are expected to dynamically lead the system from its out-of-equilibrium classical-string initial state to a thermal-equilibrium state at long times.
The relaxation dynamics toward this equilibrium state is conjectured to be initiated by a uniform, spontaneous creation of charge pairs in the bulk of the system, via a mechanism attributed to Schwinger~\cite{SchwingerMechanism}. 
A hallmark of this mechanism is the extreme sensitivity of the characteristic time scale of bulk charge-pair creation to variations in the system parameters---a manifestation of the non-perturbative origin of this effect. The observed string evolution in our experiment, nonetheless, reveals a mechanism distinct from this conventional expectation.

Explicitly, we observe charge-pair formation systematically occurring at the string edges, i.e., near the static charges, as schematically portrayed in \cref{fig_cartoon}d. This is demonstrated in the spatiotemporally resolved evolution of the charge distribution plotted in \cref{fig3: static_charge}a-i, as well as in the corresponding electric-field evolution shown in Extended Data~\cref{fig:static_charge_efield}. For vanishing or weak string tension, we find that such dynamical charge pairs perform coherent oscillations confined to the edges, as shown in \cref{fig3: static_charge}a,d,g. For larger string tension, however, these charge pairs propagate and spread from the edges towards the bulk, as shown in \cref{fig3: static_charge}c,f,i. This phenomenon---confirmed by numerical simulations shown in Extended Data~\cref{fig:static_charge_sim} and ~\cref{fig:static_charge_efield_sim}---occurs over a time scale that does not sensitively depend on Hamiltonian parameters, all the way from the region of strong to weak quantum fluctuations. 
For example, the onset and the oscillation period of pair creation at the edges remains consistent over the string-tension and coupling-strength values studied. This is in stark contrast with the exponential enhancement, as a function of string tension and coupling strength, of the rate of spatially uniform charge-pair formation in the bulk associated with the Schwinger mechanism.  We conclude that what we observe is a new, non-conventional string-breaking mechanism.

In order to elucidate the origin of this mechanism, we develop a perturbative approach, see Methods and Extended Data \cref{fig:perturbativequench} for details. This perturbative analysis demonstrates that, for small $g/J$ and $h/J$, the main contribution to non-equilibrium string dynamics arises from the quantum diffusion of a single pair of charges generated by our dynamical protocol. The associated  two-body wave function  initially peaks at the two configurations with the charge pair localized at either edge of the string---where the influence of the external static regions makes the energy cost of charge-pair creation smallest. The two-body wave function then evolves in a potential-energy landscape shaped by the coupling to the external regions and by the value of the string tension.
Our analysis reveals that, as the string tension is increased, an equipotential line opens up in configuration space, through which the charge pair propagates and spreads from the edges to the bulk.
Remarkably, our approximation captures the main qualitative features of the observed string-breaking spatiotemporal patterns in \cref{fig3: static_charge}, far away from the perturbative regime, as shown in Extended Data \cref{fig:perturbativequench}. This  analysis  provides theoretical support to our previous  conclusion that the  observed string-breaking process   differs from the Schwinger mechanism.

To confirm that the observed transient string dynamics precede the system's eventual thermalization, we compare the time evolution of a local observable with its thermal-equilibrium counterpart (see Methods) in \cref{fig3: static_charge}j-l.
We focus on the electric field, whose sign indicates the local predominance of either a string state (as in the initial state) or a vacuum state as the system evolves in time. 
For weak quantum fluctuations, \cref{fig3: static_charge}j, the string character of the state persists throughout the observed evolution, with slight variations from the motion of charges at the edges.
For stronger quantum fluctuations, \cref{fig3: static_charge}k,l, the string character is lost quicker, indicating a faster thermalization process. However, even in this regime, the system remains out of equilibrium during the entire observation time.  

\medskip

The experiments reported in this work demonstrate that quantum simulators have the 
capability to access far-from-equilibrium dynamics of interacting charges and strings in a confining gauge theory.
Our results reveal mechanisms for transient charge-formation and string-breaking dynamics, distinct from the conventional Schwinger mechanism.  
Our strategy of simulating the dynamics of charges and strings using virtual static regions can be adapted to other quantum-simulation platforms that feature individual qubit control, such as two-dimensional ion crystals~\cite{Kiesenhofer2023,guo2024} or neutral-atom arrays \cite{Chen2023,Radnaev2024,Manovitz2024}. Future experiments with fully dynamical strings and dynamical surrounding environment can elucidate the relevance of the observed string-breaking mechanism of this work, and will allow for the exploration of potential new mechanisms. Our experiments were enabled by a state-of-the-art trapped-ion architecture with fully programmable dual beam arrays. Ultimately, studies of string breaking in controlled dynamical protocols, and in particle collisions, can probe more features of string-breaking dynamics~\cite{hebenstreit2013real,hebenstreit2013simulating,kuhn2015non,pichler2016real,buyens2017real,Verdel19_ResonantSB,Magnifico2020realtimedynamics,verdel2023dynamical,Surace_2021,belyansky2024high,su2024cold,bennewitz2024simulating}. Such simulations are also within reach with programmable trapped-ion simulators~\cite{federica,de,bennewitz2024simulating}. Our strategies can also be combined with existing proposals for simulating non-Abelian and higher-dimensional lattice gauge theories~\cite{wiese2013ultracold,ZoharReview,dalmonte2016lattice,banuls2020simulating,klco2022standard,aidelsburger2022cold,bauer2023quantumsimulation,bauer2023quantum,PRXQuantum.5.037001} to enable string-breaking studies in theories of further relevance to the Standard Model.

In summary, our experimental realization of string breaking in a gauge theory marks significant progress toward the ultimate goal of first-principles quantum simulations of string fragmentation and hadronization in high-energy collisions and cosmological settings. 

\emph{Note added.} After completion of our experiments~\cite{ArinjoyTalk}, Ref.~\cite{Cochran2024} reported observation of dynamics of charges and strings in a two-dimensional array of superconducting qubits. During the preparation of this manuscript, we were also made aware of a manuscript on the observation of string breaking in a $(2+1)$-dimensional lattice gauge theory implemented in a Rydberg-atom array by QuEra Computing Inc. and collaborators.

\textbf{Acknowledgements.}  This material is based upon work supported by the U.S.~Department of Energy (DOE), Office of Science, National Quantum Information Science Research Centers, Quantum Systems Accelerator. Additional support is acknowledged from the following agencies. This work was supported in part by the National Science Foundation's (NSF's) Quantum Leap Challenge Institute for Robust Quantum Simulation (award no.~OMA-2120757). A.S., B.W., and A.V.G.~were also supported in part by the DOE ASCR Accelerated Research in Quantum Computing program (awards No.~DE-SC0020312 and No.~DE-SC0025341), AFOSR MURI, DOE ASCR Quantum Testbed Pathfinder program (awards No.~DE-SC0019040 and No.~DE-SC0024220), NSF STAQ program, and DARPA SAVaNT ADVENT. E.R.B~acknowledges support from the DOE, Office of Science, ASCR's Computational Science Graduate Fellowship (award no.~DE-SC0023112). 
A.L.~acknowledges funding through a Leverhulme-Peierls Fellowship at the University of Oxford.
F.M.S.~acknowledges support provided by the DOE, Office of Science, Office of High-Energy Physics QuantISED program, through the theory consortium ``Intersections of QIS and Theoretical Particle Physics'' at Fermilab, and by Amazon Web Services, AWS Quantum Program.
Z.D.~further acknowledges support by the DOE, Office of Science, Early Career Award (award no.~DE-SC0020271).

\textbf{Competing interests.}  All authors declare no competing interests.

\FloatBarrier

\renewcommand{\figurename}{Extended Data Fig.}
\renewcommand{\thefigure}{S\arabic{figure}}
\renewcommand{\theHfigure}{S\arabic{figure}}
\setcounter{figure}{0}

\part*{\centerline{Methods}}

\section{Theoretical Model and Analysis}

In this first part of Methods, we detail the $(1+1)$-dimensional $\mathbb{Z}_2$ lattice gauge theory that we consider and its connection with the experimentally simulated quantum-spin-chain dynamics according to Eq.~(\ref{eq_H}).
We then discuss non-equilibrium string dynamics at the lowest order in the inverse particle mass and provide a theoretical understanding of the experimentally observed charge-pair formation and spatiotemporal evolution. Lastly, we describe our calculation of thermal expectation values of the electric field, which are compared to the experimental and numerical time-evolving electric field profiles during string-breaking dynamics.

\subsection*{Mapping a $\mathbb{Z}_2$ Lattice Gauge Theory 
\\
to the Ising Spin Model}
\label{sec_mapping}
Here, we introduce a $\mathbb{Z}_2$ lattice gauge theory and illustrate how its dynamics in the gauge-invariant sector are equivalent to those of the quantum Ising chain in Eq.~(\ref{eq_H}). Our description adapts the derivation of Refs.~\cite{lerose2020quasilocalized,Surace_2021} to general variable-range Ising interactions.  
The Hamiltonian of the $(1+1)$-dimensional  $\mathbb{Z}_2$ lattice gauge theory we consider reads
\begin{align}
\label{eq_Hgauged}
    H_{\mathrm{LGT}}= &
 -g \sum_l \Big(  c^\dagger_l (b_l+b^\dagger_l) c_{l+1} +
c^\dagger_l (b_l + b^\dagger_l) c^\dagger_{l+1} + \mathrm{h.c. } 
\Big)
\nonumber\\
&
 +m \sum_l c^\dagger_l c_l 
 \nonumber\\
 & +\kappa \sum_l  n_{l} 
 -  \sum_{l }
 \sum_{ r>1}v_{r} n_{l}n_{l+r},
\end{align}
where $l\in\mathbb{Z}$ are sites of a one-dimensional lattice. In this equation, $c^\dagger_l$ and $c_l$ are 
creation and annihilation operators of fermionic particles 
on site~$l$. We call an occupied fermionic site, i.e., $c^\dagger_lc_l=1$, a  charge at site~$l$. Similarly, $b^\dagger_l$ and $b_l$ are hard-core-boson creation and annihilation operators residing on the lattice bond connecting  sites $l$ and $l+1$, with $n_{l
}=b^\dagger_{l}b_{l}$. These fields correspond to gauge-field degrees of freedom---a $\mathbb{Z}_2$ electric field.
The first line of Eq.~\eqref{eq_Hgauged} represents a minimal coupling between the matter and gauge fields, of strength~$g$; the second line is associated with the rest mass $m$ of the fermionic particles; the third line represents an electrostatic energy, including a uniform energy cost $\kappa$ for the excited electric-field configuration, as well as a variable-range self-interaction of this field, described by the couplings $v_r$.

\begin{figure*}
    \centering
    \includegraphics{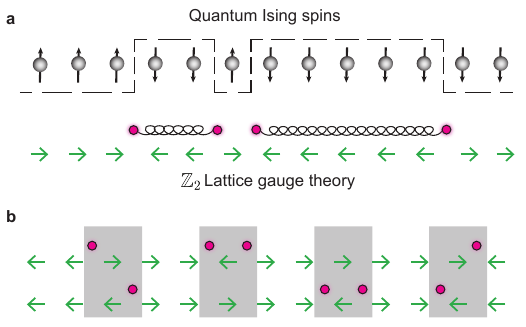}
    \caption{
    \textbf{a.} Cartoon illustration of the exact mapping between states of the quantum Ising chain in Eq.~(\ref{eq_H}) of the main text and gauge-invariant states of the $\mathbb{Z}_2$ lattice-gauge-theory Hamiltonian in Eq.~(\ref{eq_Hgauged}) satisfying Gauss's law $G_l = 1$ for all $l$.
    The mapping is based on the elimination of fermionic charge degrees of freedom via Gauss's law.
      The top row represents a classical configuration of Ising spins (simultaneous eigenstate of $\sigma^z_i$ operators for all $i$), while the bottom row represents the corresponding gauge-invariant classical configuration of matter and gauge fields (simultaneous eigenstate of the operators $c^\dagger_l c_l$ and $n_{l}$ for all $l$), with
    the red dots denoting fermionic charges (sites where $c^\dagger_l c_l=1$), and with
    horizontal green arrows encoding the two $\mathbb{Z}_2$ electric-field states (i.e., $n_l=0,1$).
    Gauss's law enforces electric-field spatial variations upon crossing lattice sites that contain a charge. This way, lattice bonds associated with an excited electric field $n_l=1$ form strings between fermionic charges. 
    \textbf{b.} The grey rectangles highlight the possible local transitions, from the top to the bottom configuration, described by the minimal-coupling term proportional to~$g$ in Eq.~(\ref{eq_Hgauged}): going from left to right, the figure shows a rightward charge hop, charge-pair annihilation, charge-pair creation, and a leftward charge hop.
    }
    \label{fig_mapping}
\end{figure*}

\begin{figure*}[t]
\includegraphics{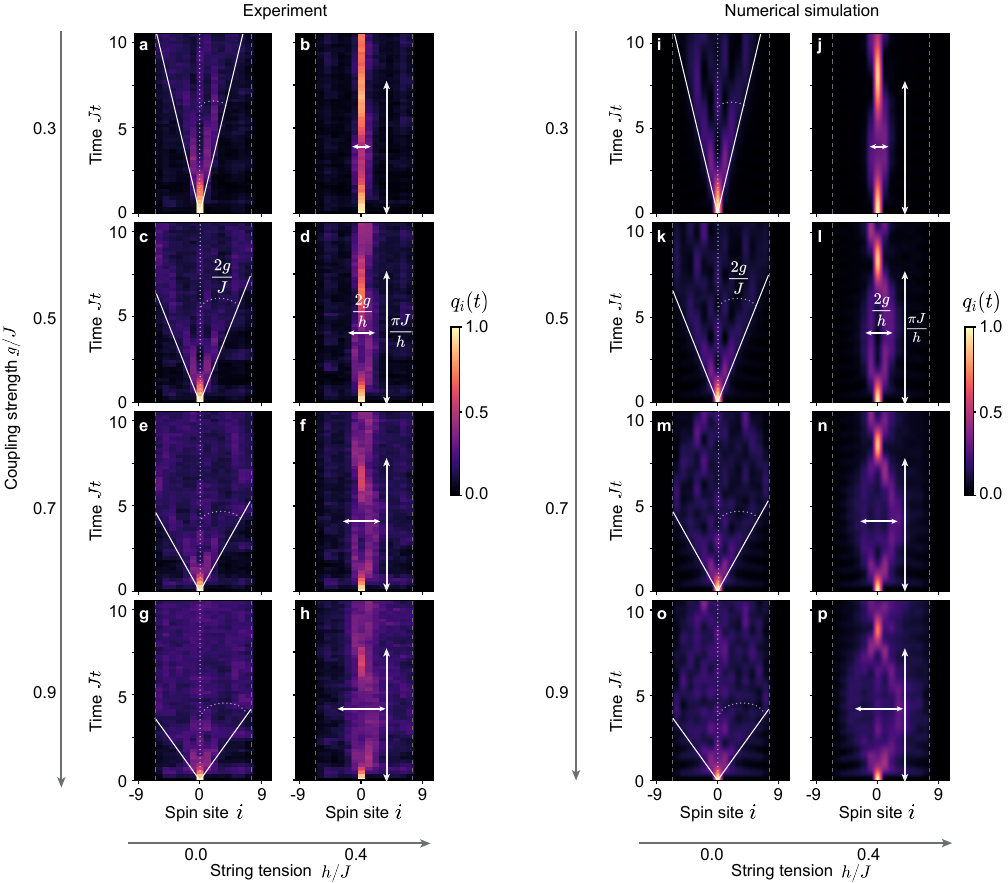}
\caption{\textbf{Comparison of experimental data and numerical simulation of isolated-charge dynamics in terms of the evolution of charge density ($q_i)$.} 
We supplement the experimental charge density data presented in \cref{fig4:localized_charge} with additional values of coupling strengths, $g/J=[0.3,0.5,0.7,0.9]$. 
The numerical simulations are performed using the experimental interaction matrices $J_{i,j}$. Plots supplement the data reported in \cref{fig4:localized_charge}.} \label{fig:domain_wall_sim_charge}
\end{figure*}

\begin{figure*}[t]
\includegraphics{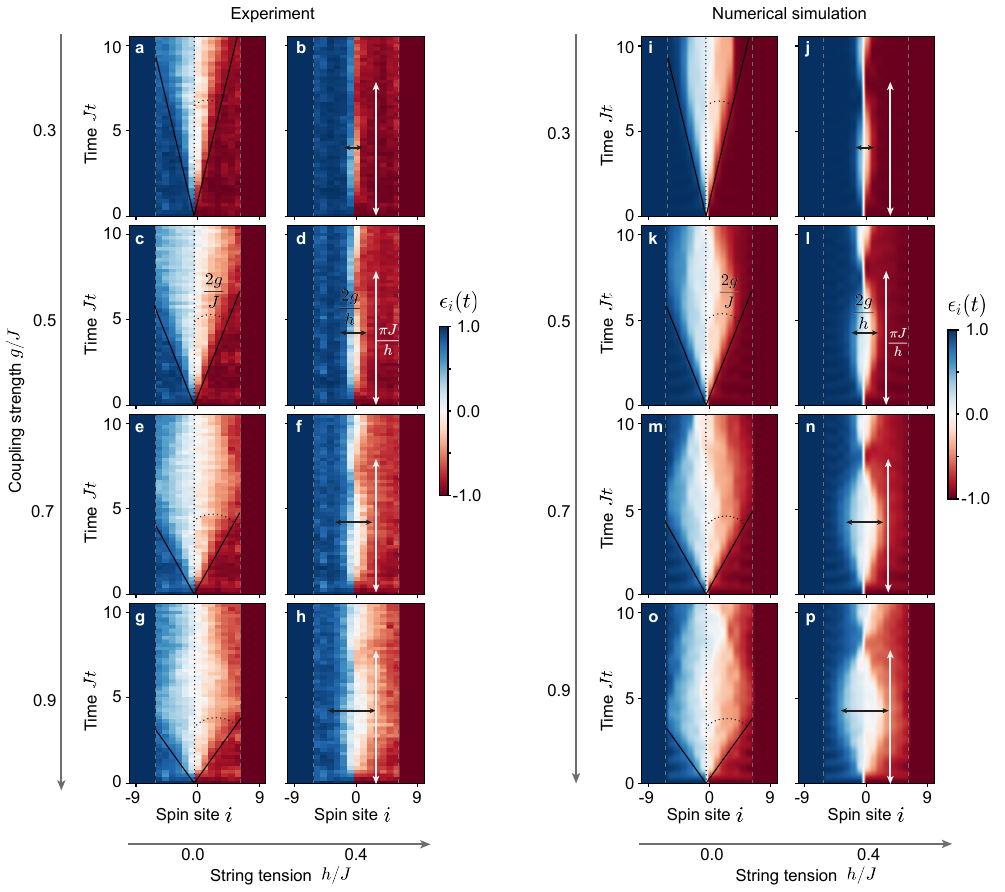}
\caption{\textbf{Comparison of experimental data and numerical simulation of isolated-charge dynamics in terms of electric-field evolution ($\epsilon_i$)}. 
We supplement the experimental electric field evolution data presented in \cref{fig4:localized_charge} with additional values of coupling strengths, $g/J=[0.3,0.5,0.7,0.9]$ 
The numerical simulations are performed using the experimental interaction matrices $J_{i,j}$. Plots supplement the data reported in \cref{fig4:localized_charge}.  \label{fig:domain_wall_sim_field}}
\end{figure*}

\begin{figure*}[t]
\includegraphics{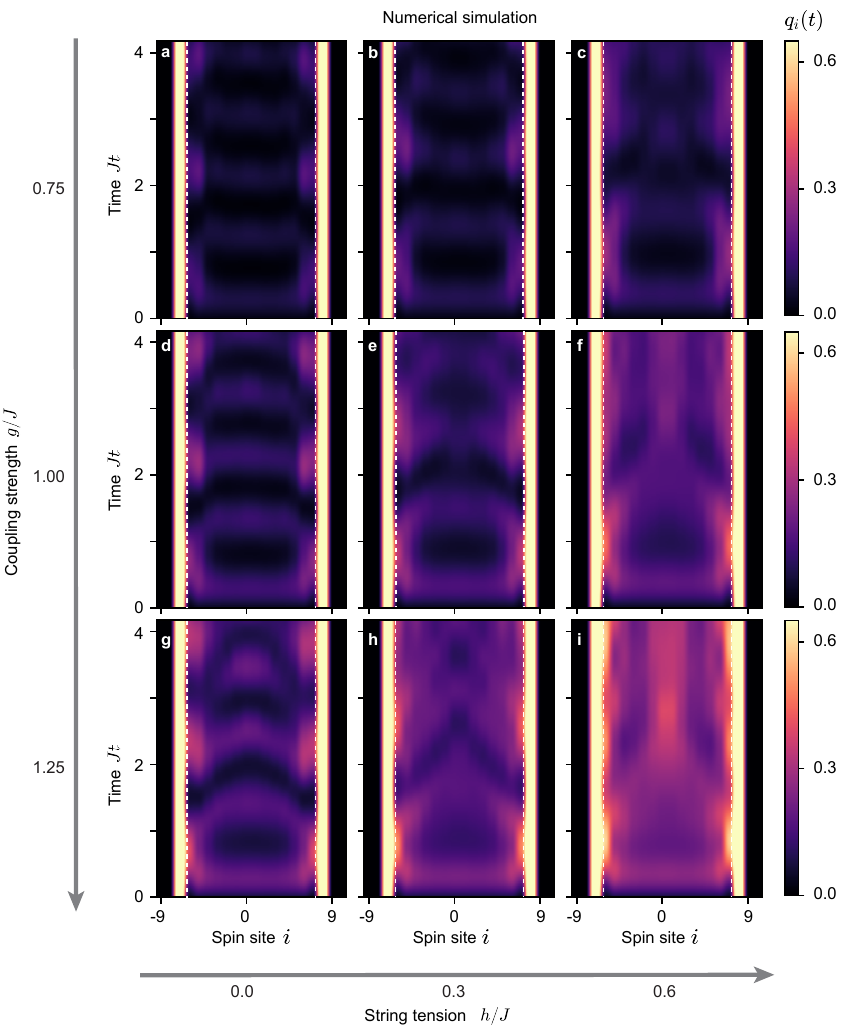}
\caption{\textbf{Numerical simulation of non-equilibrium charge dynamics}:  Numerically calculated spatiotemporal charge dynamics ($q_i$) using the same Hamiltonian parameters as in the experimental data in Fig.~\ref{fig3: static_charge}. Simulations are performed using the experimental interaction matrices. The similarity in the spatiotemporal maps between the exact numerical results and the experimental data is notable.  \label{fig:static_charge_sim}}
\end{figure*}

\begin{figure*}[t]
\includegraphics{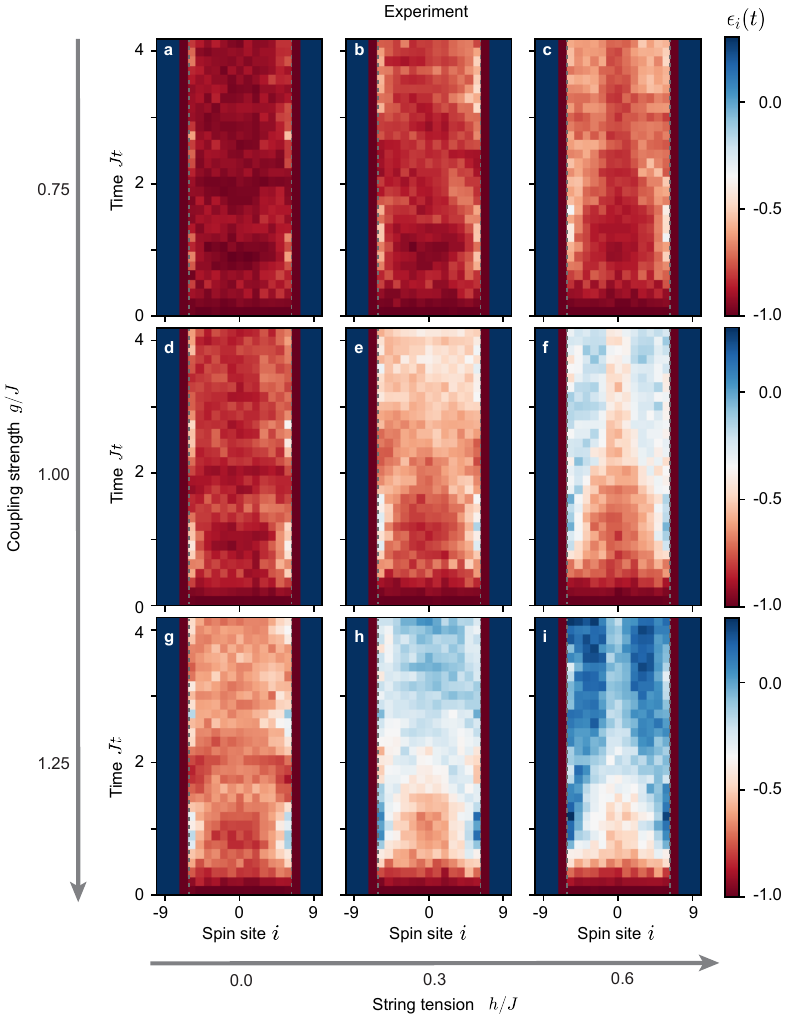}
\caption{\textbf{Experimental data on electric-field evolution ($\epsilon_i$) in non-equilibrium string dynamics.} Plots supplement the experimental charge-dynamics data reported in \cref{fig3: static_charge} with the corresponding electric-field evolution for the same experimental parameters. \label{fig:static_charge_efield}}
\end{figure*}

\begin{figure*}[t]
\includegraphics{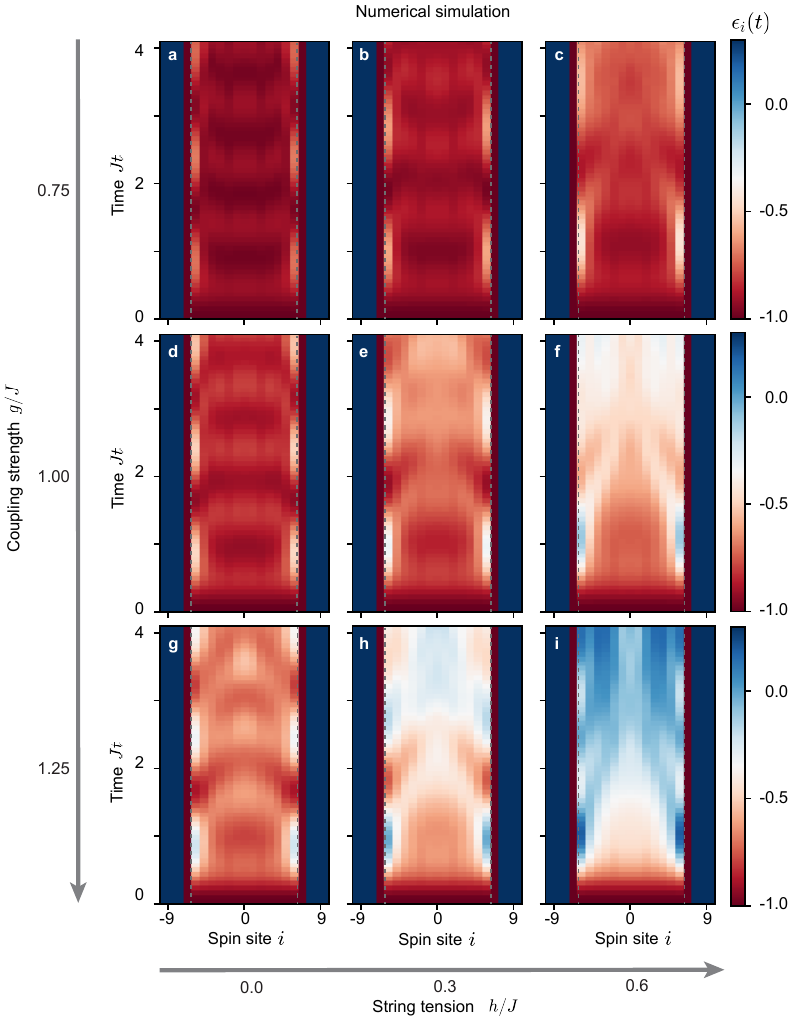}
\caption{\textbf{Numerical simulation of electric-field evolution ($\epsilon_i$) in non-equilibrium string dynamics.} Numerical simulation to be contrasted with the experimental data in the Extended Data \cref{fig:static_charge_efield}. The numerical and experimental results show good agreement. \label{fig:static_charge_efield_sim}}
\end{figure*}

The Hamiltonian in Eq.~\eqref{eq_Hgauged} is invariant under $\mathbb{Z}_2$ gauge transformations generated by the local Gauss-law operators
\begin{equation}
\label{eq_Gauss}
G_l= (-1)^{n_{l-1}+n_{l}+c^\dagger_l c_l}\, ,
\end{equation}
i.e., $[G_l,H_{\mathrm{LGT}}]=0$ for all $l$.
We will consider the gauge-invariant sector, spanned by eigenstates of the Gauss's-law operators with eigenvalue one, i.e., the sector where $G_l = 1$ for all $l$. Due to these local constraints, a configuration of the bosonic gauge fields fixes a unique configuration of the fermionic matter, see e.g., Extended Data Fig.~\ref{fig_mapping}. One can thus eliminate the fermionic degrees of freedom and obtain an exact description of the model in terms of the gauge fields only.

To formally obtain such a representation, first one applies the Jordan-Wigner transformation to turn fermions into spin-$\frac{1}{2}$ operators $\tau_l^\alpha$, for $\alpha=+,-,z$, with
\begin{equation}
    \tau_l^- = \prod_{m<l}(1-2c^\dagger_m c_m) c_l\, ,\quad \tau_l^+ = (\tau_l^-)^\dagger\, ,\quad
    \tau_l^z = 2c^\dagger_l c_l-1 \, .
\end{equation}
One can also transform hard-core bosons into spin-$\frac{1}{2}$ operators $\sigma_{l}^\alpha$, for $\alpha=+,-,z$: 
\begin{equation}
    \sigma_{l}^+ =  b_{l}, \quad \sigma^-_l = b_l^\dagger ,\quad
    \sigma_{l}^z = 1-2n_{l}.
\end{equation}
In terms of these operators, the Hamiltonian reads (up to negligible additive constants)
\begin{align}
    H_{\mathrm{LGT}}=  
    &-g \sum_l 
 \tau_l^x \, \sigma^x_{l} \,
\tau_{l+1}^x 
\nonumber\\
& + \frac m 2 \sum_l \tau_l^z 
     \, 
     \nonumber\\
     &- \frac 1 2 \bigg(
\kappa-\sum_{r>1} v_{r}\bigg)  \sum_l  \sigma^z_{l} 
- \frac 1 4  \sum_{l}
     \sum_{ r>1} v_{r} \sigma^z_{l}\sigma^z_{l+r},
\end{align}
with the constraint $
    G_l=-\sigma_{l-1}^z \tau_l^z \sigma_{l}^z =1$.
We now introduce a unitary transformation $U$ such that the transformed Gauss's law $G_l'=UG_lU^\dagger = 1$  only depends on the matter degrees of freedom, whereas the transformed Hamiltonian $H_{\mathrm{LGT}}'=UH_{\mathrm{LGT}}U^\dagger$ only depends on the gauge-field degrees of freedom. 
This is accomplished by choosing
\begin{equation}
    U=\prod_l \exp\left[\frac{i\pi}{2}(\tau_{l}^x-1)\frac{1-\sigma_{l-1}^z\sigma_{l}^z}{2}\right] \, .
\end{equation}
The transformed constraint
$
    G_l'= -\tau_l^z = 1
$
decouples the $\tau$ spins. Subsequently, the transformed Hamiltonian depends only on the $\sigma$ spins: 
\begin{align}
\label{eq_transfH}
    H_{\mathrm{LGT}}'= & -g \sum_l 
 \sigma^x_{l} 
     \, 
     \nonumber\\
     & + m \sum_l \frac{1-\sigma_{l-1}^z  \sigma_{l}^z}{2} \,
     \nonumber\\
     &-\frac 1 2 \bigg(
 \kappa-\sum_{r>1} v_{r}\bigg)  \sum_l  \sigma^z_{l} 
 - \frac 1 4  \sum_{l}
     \sum_{ r>1}v_{r} \sigma^z_{l}\sigma^z_{l+r}.
\end{align}
We now identify the bonds of the fermionic chain connecting sites $l$ and $l+1$ with sites $i$ of the dual spin chain, and the parameters as
\begin{equation}
\label{eq_parameters}
    m=2J_1, \quad 
    \kappa = 2h+ \sum_{r=2}^\infty J_r, \quad v_r =  4 J_{r} \, .
\end{equation}
Here, $J_r \equiv J_{i,i+r}$ is the spatial profile of translationally invariant Ising interactions. One can then readily observe that Eq.~(\ref{eq_transfH}) coincides (up to an irrelevant additive constant) with the Hamiltonian in Eq.~(\ref{eq_H}) implemented by our trapped-ion simulator.

\subsection*{Perturbative Analysis of String Dynamics}
Here, we provide theoretical understanding of non-equilibrium string-breaking dynamics in the limit where the particle mass is the dominant energy scale,  
$g,h,J_2,J_3,J_4,\dots\ll J_1 = J$, see Eq.~\eqref{eq_parameters}.  
In this regime, the evolution of the classical-string  initial state $|\Psi_0\rangle$ (depicted in Fig.~\ref{fig_cartoon}c) is accurately described for times $0<t \ll J/g^2$ by a perturbative expression of the form
\begin{equation}
|\Psi(t)\rangle = e^{-iE_0 t} \Big[ |\Psi_0\rangle +
\Big( |\Psi_1(t)\rangle-|\Psi_1(0)\rangle \Big)  + \mathcal{O}\big(g^2 t/J\big) \Big]  ,
\end{equation}
where $E_0=\langle \Psi_0 | H | \Psi_0 \rangle$ is the unperturbed string energy, and the leading-time non-trivial dependence, $|\Psi_1(t)\rangle$, describes the spreading dynamics of a \textit{conserved} number of charge pairs generated by the abrupt perturbation at $t=0$.

In order to calculate $|\Psi_1(t)\rangle$, let us split the infinite-spin-chain Hamiltonian as $H=H_0+V_{\rm{diag}}+V_{\rm{offdiag}}$, where the dominant part $H_0=- J \sum_i \sigma^z_i\sigma^z_{i+1} $  
defines subspaces with different numbers of charges; the remainder is split into a block-diagonal component $V_{\rm{diag}}= - \sum_i \sum_{r\ge 2} J_r \sigma^z_i\sigma^z_{i+r} - h \sum_i \sigma^z_i - g \sum_{i_0\le i \le i_0+L-1} \big(P_{i-1}^{\uparrow} \sigma^x_i P_{i+1}^{\downarrow}+P_{i-1}^{\downarrow} \sigma^x_i P_{i+1}^{\uparrow}\big)$, which conserves the number of charges (i.e., $[H_0,V_{\rm{diag}}]=0$), and an off-block-diagonal component $V_{\rm{offdiag}}= - g \sum_{i_0\le i \le i_0+L-1} \big(P_{i-1}^{\uparrow}\sigma^x_i P_{i+1}^{\uparrow}+P_{i-1}^{\downarrow}\sigma^x_i P_{i+1}^{\downarrow}\big)$,  which creates or destroys pairs of charges [here $P^{\uparrow}_i$ ($P^{\downarrow}_i$) projects spin $i$ on state $\ket{\uparrow}$ ($\ket{\downarrow}$)].
Notice that the transverse-field perturbation only acts on the spins in the simulated physical region $[i_0,i_0+L-1]$; in our experiments we have $L=13$ and choose a labelling of spin sites such that $i_0=-(L-1)/2=-6$.

We seek a unitary that 
transforms the Hamiltonian into a block-diagonal form, order by order in perturbation theory. At the lowest order, a transformation of the form $e^{iS}$ is sought such that $e^{iS} H e^{-iS}= E_0+ H_{\rm{eff}} + \mathcal{O}(g^2/J)$, where $[H_{\rm{eff}},H_0]=0$. This requires choosing a block-off-diagonal generator $S$ satisfying $i[S,H_0+V_{\rm{diag}}]=-V_{\rm{offdiag}}$.
By computing the time evolution of the transformed initial state $ e^{iS} |\Psi_0\rangle$ under the transformed Hamiltonian,  we obtain a perturbative expansion of the time-evolving state,
\begin{equation}
\label{eq_perturbative}
|\Psi(t)\rangle = e^{-itE_0} \bigg[ \mathbbm{1}+i
\Big( e^{-it H_{\rm{eff}}}-\mathbbm{1}\Big) S + \mathcal{O}\big(g^2 t/J\big) \bigg] |\Psi_0\rangle \, .
\end{equation}
From Eq.~\eqref{eq_perturbative}, we read off 
\begin{equation}
\label{eq_psi1}|\Psi_1(t)\rangle = i e^{-it H_{\rm{eff}}} S |\Psi_0\rangle.
\end{equation}
This procedure can be iterated to arbitrarily high orders in perturbation theory, to find an accurate description of time evolution over an increasingly long time window. We will show that the lowest-order analysis is sufficient to grasp the qualitative features observed in our experiment.

The picture of string dynamics resulting from this time-dependent perturbation theory has a transparent physical interpretation: 1) The abrupt perturbation at time $t=0$ generates a quantum superposition of charge pairs at different locations, locally breaking the string; 2) Subsequently, these charges undergo quantum-mechanical motion across the physical region during time evolution. Processes 1) and 2) are encoded in the leading time dependence $|\Psi_1(t)\rangle$ in Eq.~\eqref{eq_psi1}. Specifically, process 1) is  described by the action of the operator $S$---which is by construction an off-block-diagonal operator---on the classical string state $|\Psi_0\rangle$. At lowest order, $S$ flips a single spin, i.e., it creates a single pair of charges adjacent to each other. The amplitude of this process is affected by the position-dependent energy cost of creating a pair, resulting in a spatially inhomogeneous initial wave function of the charge pair (see below for its expression). Process 2) is described by the time-evolution operator $e^{-it H_{\rm{eff}}}$   associated with the number-conserving effective Hamiltonian $H_{\rm{eff}}$.
These dynamics are affected by energy of the two-charge configurations, which depend on the two positions, resulting in a quantum diffusion through an inhomogeneous potential landscape (see below for its expression).
 
At the lowest order in perturbation theory, string dynamics thus reduce to a two-body problem (for the two charges created), which can be straightforwardly analyzed. 
Its solution provides key insights into string-breaking dynamics, and determines the time-dependent probability $\mathcal{P}(t)$ of finding the system in a broken-string configuration,
\begin{equation}
\mathcal{P}(t)= 2  \big(\langle \Psi_1(0)| \Psi_1(0) \rangle - \text{Re} \langle \Psi_1(0)| \Psi_1(t) \rangle \big) + \mathcal{O}\big((g^2t/J)^2\big)
\end{equation}
(notice that the norm of $\ket{\Psi_1(t)}$ is perturbatively small).

Let us now analyze the two-body problem. We denote by $|l_1,l_2\rangle$ the configuration with the two charges sitting at lattice sites $l_1$ and $l_2$, with $i_0\le l_1 < l_2 \le i_0+L$ (corresponding to the $L+1$ spin bonds involving at least one dynamical spin). 
Considering an exponentially decaying interaction profile $J_{r}={J} e^{-\beta(r-1)}$, the energy of the charge-pair configuration relative to the intact classical string, i.e., $V_{l_1,l_2}=
\langle l_1,l_2 | H | l_1,l_2 \rangle - E_0$, is given by
\begin{align}
\label{eq_perturbativeconfigurationalenergy}
V_{l_1,l_2} =
\frac {4J} {(1-e^{-\beta})^2}\Big(&
1+e^{-\beta l_2}-e^{-\beta l_1} - e^{-\beta (l_2-l_1)}
\nonumber\\
&+ e^{-\beta (L+2-l_1)}-e^{-\beta (L+2-l_2)}
\Big)
\nonumber\\
&-2h(l_2-l_1).
\end{align}
The initial state of the two charges can then be expressed as
\begin{equation}
\label{eq_perturbativeinitialstate}
|\Psi_1(0)\rangle = i S |\Psi_0\rangle=i \sum_{l=i_0}^{i_0+L-1} \frac {g} {V_{l,l+1}} |l,l+1\rangle,
\end{equation}
and the effective Hamiltonian reads
\begin{align}
&H_{\rm{eff}}=\sum_{l_1 <l_2}  V_{l_1,l_2}\, |l_1,l_2\rangle\langle l_1,l_2| 
\nonumber\\
&~~- g 
\sum_{l_1 <l_2} \Big(
|l_1-1,l_2\rangle\langle l_1,l_2|
+|l_1,l_2+1\rangle\langle l_1,l_2|+ \rm{h.c.}\Big)
.
\end{align}
Dynamics governed by this Hamiltonian can be viewed as a single particle hopping in a two-dimensional lattice of triangular shape with uniform amplitude $g$ and hard-wall boundaries, under the influence of an inhomogeneous potential given by Eq.~\eqref{eq_perturbativeconfigurationalenergy}.
Higher-order terms contain corrections to the potential and longer-range hopping processes.

\begin{figure*}
\centering
\includegraphics{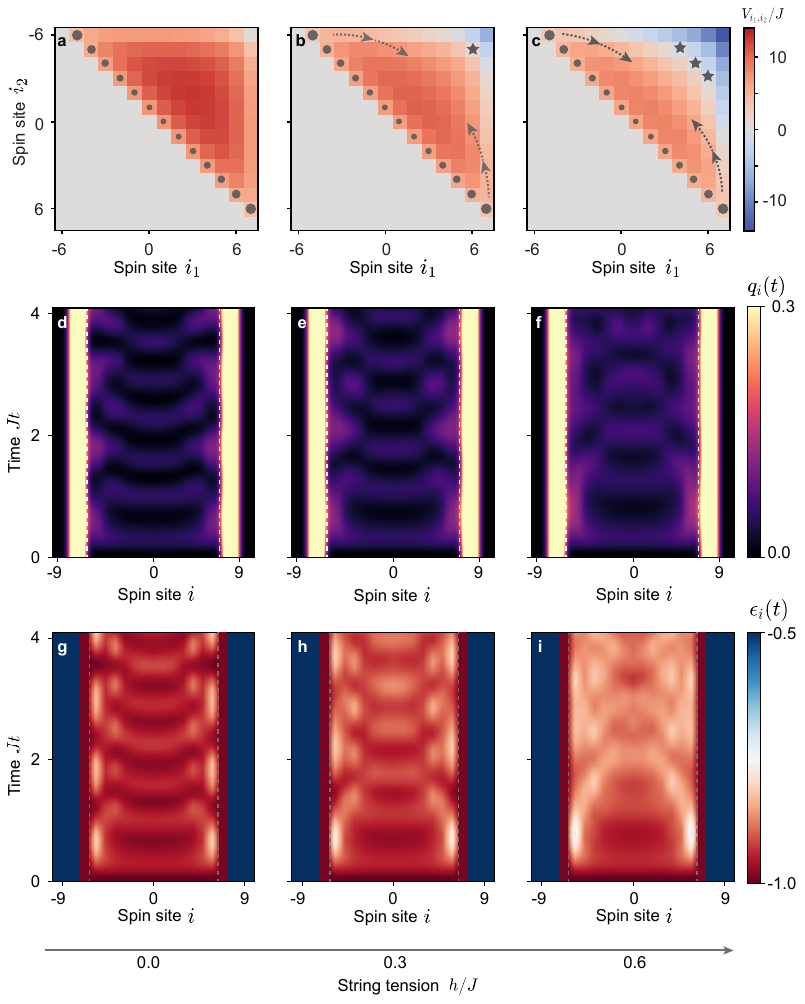}
\caption{\textbf{Effective two-charge description of non-equilibrium string dynamics.} 
\textbf{a-c.} 
Color maps of the effective two-body potential $V_{l_1,l_2}$ in Eq.~\eqref{eq_perturbativeconfigurationalenergy} for exponentially decaying Ising interactions with exponent $\beta=0.78$ and for the three values of $h/J$ indicated, as in our experimental simulations. The two coordinates represent the positions of the two charges.
The size of the shaded dots is proportional to the initial two-body probability distribution $|\langle l_1,l_2|\Psi_1(0)\rangle|^2$, see Eq.~\eqref{eq_perturbativeinitialstate}. The shaded arrows highlight approximate equipotential lines passing through the two edge configurations. Quantum spreading dynamics across this potential-energy landscape underlies the string-breaking spatiotemporal patterns  propagating from the edges into the bulk, which we observed in the experiments for sufficiently strong string tension. 
The black stars indicate resonant charge-pair configurations, which can be non-perturbatively generated over long times via the conventional Schwinger mechanism. \textbf{d-i.} Spatiotemporal charge evolution (panels \textbf{d-f}) and the corresponding electric-field evolution (panels \textbf{g-i}) within the perturbative approach expressed by Eq.~\eqref{eq_perturbative}, for $g/J=0.75$ and the $h/J$ values indicated. Despite these parameters being far from the perturbative regime where quantitative accuracy would be expected, the qualitative similarity with experimental data is remarkable. 
}
\label{fig:perturbativequench}
\end{figure*}

Extended Data~\cref{fig:perturbativequench}a-c illustrates this description of charge dynamics, with a color-map of the potential $V_{l_1,l_2}$. 
The appearance of negative values of the potential in the top-right corner for sufficiently large $h$ reflects \emph{static} string breaking, i.e., a change from a string-like to a broken-string-like ground state.
The initial state is represented by a wave function supported along the diagonal side of the domain (shaded dots), with amplitude peaks at the two edges. From these initial positions, the particle may then quantum-mechanically propagate through the landscape. In a semiclassical description---realized in an appropriate limit of long string $L\to\infty$ and smoothly decaying interactions with $\beta\to0$---the particle would be constrained to propagate along equipotential lines. Superimposed arrows in Extended Data~\cref{fig:perturbativequench}b-c highlight approximate equipotential lines, which appear as the string tension $h$ is increased, thereby opening up a \emph{channel} for charge pairs generated at the edges to spread into the bulk. 
This mechanism underlies the qualitative features of the spatiotemporal string-breaking patterns observed in our experiment, see \cref{fig3: static_charge}.

Instances of spatiotemporal charge and electric-field evolution
as approximated by the two-body time-dependent Schr\"{o}dinger equation are shown in Extended Data Fig.~\ref{fig:perturbativequench}d-i for several values of $h/J$ and the lowest value $g/J=0.75$ for the coupling strength in our experimental simulations. These parameters are far from the theoretical perturbative limit, where the approximation would be quantitatively accurate. Nonetheless, the dynamical behavior exhibits
qualitative similarity with the exact string dynamics, 
approximately capturing, in particular, the observed dynamical string-breaking mechanism.

Long-time string dynamics, far beyond the transient scale $0<t\ll J/g^2$, are expected to be described by the Schwinger mechanism---a mechanism associated with the spontaneous materialization of resonant charge pairs. The spatiotemporal rate of this non-perturbative bulk process is suppressed faster than any power of~$g$~\cite{SchwingerMechanism,hebenstreit2013real,lerose2020quasilocalized}, precluding observation over the accessible time scales in our experiment. Within the picture in Extended Data~\cref{fig:perturbativequench}a-c, resonant charge-pair configurations satisfy $V_{l_1,l_2}\approx 0$ and are indicated in the plots by black stars.

\subsection*{Thermal Expectation Values}

Here, we outline the calculation of thermal expectation values of the electric field in the main text, Fig.~\ref{fig3: static_charge}. The state of a quantum many-body system far from equilibrium is described by a superposition of a large number of high-energy eigenstates of the system's Hamiltonian.
In generic interacting systems, local observables are expected to relax to stationary values compatible with a thermal-ensemble average. The corresponding thermal-equilibrium state is given by the canonical Gibbs ensemble, $\rho_{\rm Gibbs}(T) = e^{-H/(k_BT)}/\Tr{\left[ e^{-H/(k_BT)} \right]}$, with Boltzmann factor $k_B$ and with the temperature $T$ fixed by the total energy. For each of the systems analyzed in \cref{fig3: static_charge}, the temperature is determined by solving $\bra{
\Psi_0} H \ket{
\Psi_0} = \Tr \left[ H \rho_{\rm Gibbs}(T) \right] $, where $\ket{\Psi_0}$ is the classical-string state depicted in Fig.~\ref{fig_cartoon}c. The thermal-equilibrium profiles for the 
electric field displayed in \cref{fig3: static_charge}j-l are computed via $\langle \sigma_z^i \rangle_{\rm thermal} = \Tr[\sigma_z^i \rho_{\rm Gibbs}(T)]$.
\\

\section{Experimental Apparatus and Hamiltonian Engineering
}
In this section, we first introduce the basic experimental features of the trapped-ion apparatus employed in this work. We then describe the methods used to generate the programmable spin-spin interaction and the 3D magnetic field used to engineer the Hamiltonian in \cref{eq_H}.

\subsection*{Trapped-ion apparatus}
The trapped-ion apparatus employed in this work features a linear array of $L=15\quad ^{171}$Yb$^{+}$ ions trapped in a microfabricated surface ion trap \cite{schuckert2023,katz2024observing,Feng2023}. We encode the spin states in the two clock levels of the ${^2S_{1/2}}$ ground state, where ${\ket{\uparrow}_x\equiv\ket{F=1, m_F=0}}$ and ${\ket{\downarrow}_x\equiv\ket{F=0, m_F=0}}$ with an energy separation of ${\omega_0=2\pi\times12.6}$ GHz. At the beginning of each experimental cycle, the ions are cooled using Doppler cooling. The transverse motional modes used to generate long-range interactions are further cooled down to their motional ground state using resolved-sideband cooling. 
The initial state is prepared by optically pumping the ions to $\ket{\downarrow\cdots\downarrow}_x$ state, and rotating them to the $z$-basis with $R_{y}(- \pi/2)$ gates, i.e., single-qubit rotations around the $y$ axis of the Bloch sphere by angle $- \pi/2$. Such gates are implemented by tuning the Raman beatnote to drive the carrier transition. After the state preparation, a spin-spin interaction is generated to realize the Hamiltonian in Eq.~\eqref{eq_H} (see the following subsections), and the spins evolve under this Hamiltonian for a variable duration. We measure the final spin-state population along the $z$ basis by applying a $R_{y}(-\pi/2)$ gate, followed by spin-dependent fluorescence detection. 

The optical setup consists of two global beams and a dual array of tightly focused laser beams, each with independently controllable optical frequency, phase, and amplitude. All beams originate from a single optical frequency comb pulsed laser at 355nm. The global beams are generated by the passage of one beam through an acousto-optic modulator (AOM), which splits the beam into two spatially and frequency-distinct components by shifting the radio-frequency (RF) beatnotes by $\omega^\text{glob}_0\pm(\omega_L+\mu)$. $\omega_L$ is the frequency of the lowest-frequency radial phonon mode, the \emph{zig-zag} mode, and $\mu$ is the detuning of the Raman beatnote relative to this phonon mode. These two components are combined and directed onto the ions via a telecentric optical path, ensuring uniform illumination across the ion chain.

The dual beam arrays originate from a single beam that passes through a diffractive optical element, initially generating an array of more than thirty tightly focused laser beams. Each beam then passes through an independent channel of a multichannel AOM, which applies two RF frequencies, generating two first-order diffracted beams that are spatially separated, with controllable frequencies, amplitudes, and phases. The optical setup is precisely designed and aligned to overlap the two distinct beam arrays onto the ion plane, ensuring that each pair of tightly focused laser beams addresses the ions perpendicular to the combined global beam. The system is stabilized using an active vibration control platform to minimize the sensitivity of the optical phase to acoustic vibrations. We routinely calibrated the relative phase at the ion positions to account for any changes or drifts in the optical phase between the pairs of beams in the dual array, due to differences in their optical paths. Additionally, by controlling the radial trapping potential, we aligned the wavevector difference between the (combined) global beam and the dual beam array so that only a single set out of $L$ phonon modes, associated with a specific direction of motion, is addressed by the beams.

\subsection*{Generation of Spin-Spin Interaction}
The spin-spin Ising interaction in the trapped-ion quantum simulator is generated by applying spin-dependent forces, achieved by simultaneously driving the red and blue motional sidebands. This is facilitated by the two components of the global beams and one array of tightly focused laser beams. The second array 
is used to generate effective magnetic fields, which will be described in the following subsections. The multichannel AOM shifts the frequency of each beam in the first array by $\omega^\text{ind}_0$, which satisfies the condition $|\omega^\text{glob}_0-\omega^\text{ind}_0|=\omega_0$. Consequently, the Raman beatnotes $\omega_0\pm(\omega_L+\mu)$ of the global beam, along with the first array, simultaneously drive a blue and red motional sideband transition, generating an effective spin-spin Ising interaction  $H_{ZZ}=\sum_{i,j}J_{i,j}\sigma_i^z\sigma_j^z$ via the M\o lmer-S\o rensen protocol~\cite{Monroe2021}. The experimental $J_{i,j}$ matrix for $L$ spins 
has the form~\cite{Monroe2021, Feng2023},
\begin{equation}
J_{i,j}= \sum_{k=1}^L\frac{ \eta_{i,k} \eta_{j,k}\Omega_i\Omega_j}{\omega_L+\mu-\omega_k}\label{eqn:Jij},
\end{equation}
where $\eta_{i,k}=0.08 b_{i,k}$ are the Lamb-Dicke parameters describing the coupling between spin $i$ and motional mode $k$ with $b_{i,k}$ being the mode-participation matrix elements. $\Omega_i$ is the carrier Rabi frequency at ion $i$, and $\omega_k$ is the radial motional frequency of mode $k$, labelled in decreasing order with $1\leq k\leq L$. 

By precisely tuning the axial trap potential~\cite{schuckert2023, Feng2023}, we achieve a nearly uniform ion spacing of $3.75~\mu\text{m}$ for the 13 central ions in the 15-ion chain. In this experiment, we control the phase and amplitude of the 13 individual beams of the first array via the RF signals driving the corresponding AOM channels, while keeping the remaining beams off. Therefore, the two edge ions assist in trapping and participate in the motion, but they are not driven by the Raman interaction, and their spins do not contribute to the Hamiltonian in Eq.~\eqref{eq_H}.  This capability allows for varying the system size without changing the trapping potential~\cite{Feng2023}, which might otherwise alter the interaction matrix~\cite{de2023}. 

One can numerically calculate the interaction matrix elements $J_{i,j}$ in \cref{eqn:Jij} using the axial trapping parameters. We find that these elements approximately decay with a functional form best described by $J_r = Je^{-\beta(r-1)}r^{-\alpha}$, where $r = |i-j|$, and the average interaction strength is $J_r = \frac{1}{L-r} \sum_i \left|J_{i,i+r}\right|$. The range of interaction can be adjusted by changing the symmetric detuning $|\mu|$ of the global beam relative to the red and blue sideband---for large detuning, the exponential term dominates at short distances, and for small detuning, the interaction becomes longer-range~\cite{Feng2023}. For our chosen trap parameters, the zig-zag mode frequency is $\omega_L = 2\pi \times 2.78$~MHz, and the detuning is $\mu = -2\pi \times 35$~kHz. For this configuration, we find that the numerically calculated interaction matrix follows the decay pattern of $J_r$ with $\alpha = 0$ and $\beta = 0.78$. 

The participation matrix elements ($b_{i,L}$) of the zig-zag mode have alternating signs for even and odd sites, which causes the $J_{i,j}$ matrix elements to be staggered, i.e. $\text{sgn}\left(J_{r}\right)=-\text{sgn}\left(J_{r+1}\right)$.
To correct this form of the interaction and ensure that all the $J_{i,j}$ elements have the same sign, we shift the optical phase of every other beam in the first array by $\pi$ while generating the spin-spin interaction.

\begin{figure}[t]
\includegraphics{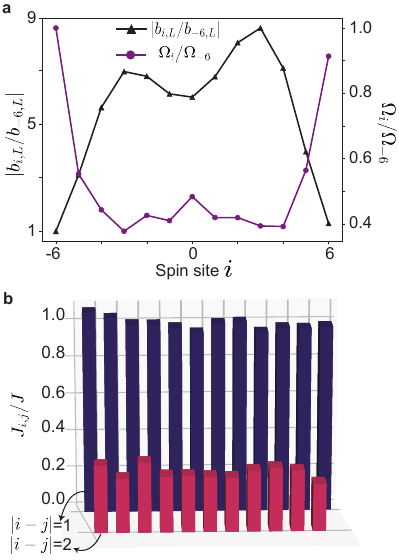}
\caption{\textbf{Programmability of the spin-spin interaction.} \textbf{a.} Absolute values of the mode-participation matrix elements, $b_{i,k}$, for the lowest-frequency mode ($k=L$) are shown in black triangles. The matrix elements are normalized by the participation factor of the left edge ion, $b_{-6,L}$. Ions at the center of the chain participate more in this mode. Since we operate at a detuning that is closer to the lowest-frequency mode than any other motional frequency, the interaction profile across the chain is mainly governed by the participation of this mode, which renders the interaction profile non-uniform across the chain [see (Eq.~\ref{eqn:Jij})]. In order to make the interaction profile uniform, we design a non-uniform amplitude profile of the individual beam array ($\Omega_i$), as shown in purple circles. For ease of representation, we normalize the amplitudes with that of the beam addressing the left edge ion, $\Omega_{-6}$. \textbf{b.} The experimental measurement of approximately uniform nearest-neighbor (NN) and next-nearest-neighbor (NNN) coupling strengths in a $L=13$ ion chain. We applied the amplitude profile shown in \textbf{a}, up to an overall scale factor, to obtain the uniform profile with an average NN strength $J=2\pi\times 0.34$ kHz. For simplicity of presentation, we have normalized the NN and NNN values with $J$.  \label{fig:Jij_calibration}}
\end{figure}

The amplitudes of the mode-participation matrix of the zig-zag mode are non-uniform (black triangles in Extended Data Fig.~\ref{fig:Jij_calibration}a), leading to a non-uniform interaction matrix across the chain. To correct for this non-uniformity, we adjust the amplitudes of the individual beams by controlling the RF amplitude driving each AOM channel. For a chain of $L$ spins, we experimentally measure $L-1$ nearest-neighbor (NN) and $L-2$ next-nearest-neighbor (NNN) couplings. By tuning the individual beam amplitudes, we achieve approximately uniform NN couplings and minimize fluctuations in the NNN couplings. The Rabi frequencies of the individual beams are represented by purple dots in Extended Data Fig.~\ref{fig:Jij_calibration}a. A comparison of the mode-participation factors with the beam-amplitude profiles indicates that the amplitude profile is the inverse of the participation factors. After applying this amplitude profile, up to a scaling factor, we experimentally observe near uniform NN and NNN couplings as shown in Extended Data Fig.~\ref{fig:Jij_calibration}b. Across the chain, the average NN coupling reads $J=J_1 = 2\pi \times 0.34(1)$~kHz, and NNN coupling reads $J_2 = 2\pi \times 0.10(2)$~kHz, where the error indicates the standard deviation across the chain. 

When the spin-spin interaction is quickly turned on, an off-resonant coupling to the carrier transition can be detected. To avoid this undesired coupling, the individual beam amplitudes are linearly ramped on with a 5$\mu$s duration. This ramp time is significantly longer than the off-resonant carrier oscillation time scale $1/\sqrt{\Omega_{i}^2+(\omega_{L}+\mu)^2}\approx1/(\omega_{L}+\mu)$, but remains much shorter than $1/J_1$. As a result, the spin-spin interaction is still effectively turned on instantaneously. 

\subsection*{Generation of Transverse Magnetic Field}
In addition to the Ising interaction, an effective local transverse field is applied by adjusting the frequencies of individual beams. Specifically, if the optical beatnote frequencies addressing ion $i$ are shifted by $g_i$ relative to the qubit resonance, this detuning induces an effective static local transverse field term $g_i\sigma_i^x$ in the Hamiltonian, under the condition $g_i\lesssim J$~\cite{Aaron2016,Feng2023}. We realize this term by shifting the RF frequencies of the dual-array beams. The use of inhomogeneous beam amplitudes, as discussed earlier, inevitably results in an inhomogeneous AC Stark shift across the ion chain. These shifts manifest as unwanted $\epsilon_i\sigma_i^x$ terms in the interaction Hamiltonian. We experimentally measure these shifts and apply an equal but opposite frequency shift for each tightly focused beam in the dual arrays, effectively generating Hamiltonian terms $-\epsilon_i\sigma_i^x$, which exactly cancel the Stark shift effects and ensure a uniform transverse field across the chain.

Due to the slight misalignment of the spatial overlap between the two frequency components of the global beam, and their finite sizes, the power imbalance between these components results in a large Stark shift error on the edge ions. To mitigate this, we implemented dynamical-decoupling pulses on the edge spins in the $13$-spin chain, minimizing their response to shot-to-shot fluctuations in the Stark shifts~\cite{Morong2022}.

\medskip

\subsection*{Generation of Longitudinal Magnetic Field}
The dual-array setup enables the implementation of fully programmable local longitudinal fields, applied simultaneously with the transverse-field Ising Hamiltonian terms. To achieve this, we introduce additional frequency shifts of $\omega^{\text{ind},i}_0+\omega_L+\mu$ in the multichannel AOM, generating a second array of tightly focused beams. Through a telecentric beam path, they are overlapped with the beams generating the spin-spin interactions, as shown in Fig. \ref{fig_cartoon}e. The second array of beams, along with the $\omega^{\text{glob}}_0+\omega_L+\mu$ frequency component of the global beam, produces a beatnote that drives the qubit transition on resonance at frequency $\omega_0$. This is equivalent to a site-dependent magnetic field \(h_i^\phi(t)\sigma_i^\phi\), where \(\sigma_i^\phi=\cos\phi \, \sigma_i^z+\sin\phi \, \sigma_i^y\), enabling control of all three components of the local magnetic field of the $i$th spin. 

For each individual beam, we calibrate the Rabi frequency versus the RF amplitude on the AOM, and additionally shift the phase of the RF signal to align the field in the $z$ axis (owing to a difference in the optical path relative to the first array) such that \(\sigma_{i}^\phi=\sigma_{i}^z\). This allows us to apply the tunable inhomogeneous longitudinal-field profiles necessary for the experiments.

\begin{figure}[t]
\includegraphics{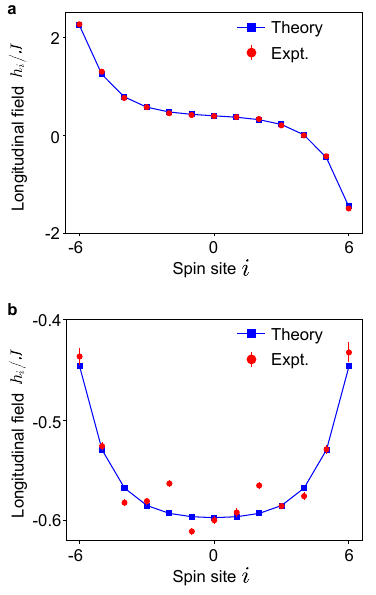}
\caption{\textbf{Examples of inhomogeneous longitudinal-field profiles.} The comparison of theoretical and measured longitudinal fields for the different configurations explored in this work. \textbf{a.} Inhomogeneous profile in the presence of an isolated charge,  corresponding to a homogeneous field of $h/J=0.4$. \textbf{b.} Inhomogeneous profile corresponding to $h/J=0.6$ in the presence of virtual static charges. The error bars represent fitting errors during experimental calibration. (The error bars are smaller than the dots in \textbf{a}.) \label{fig:Bx_calibration}}
\end{figure}

In this work, we use two different configurations of virtual static spins for studying isolated-charge dynamics (\cref{fig_cartoon}b) and dynamical string breaking (\cref{fig_cartoon}c,d). The inhomogeneous fields corresponding to these configurations are, respectively, given by
\begin{align}
    \Delta h_i^{(\rm{charge})}=&   \sum_{r=i}^\infty J_r - \sum_{r=L+1-i}^\infty J_r, \\
    \Delta h_i^{(\rm{string})} =&  -J_i + \sum_{r=i+1}^\infty J_r -J_{L+1-i} + \sum_{r=L+2-i}^\infty J_r,
\end{align}
with $J_r = J e^{-0.78(r-1)}$ in our specific setup.
The numerical and measured values associated with the experimental parameters are reported in Extended Data \cref{fig:Bx_calibration}.

\clearpage
\bibliography{Refs}

\end{document}